\documentclass[pre,aps,showpacs,twocolumn,amsmath,letterpaper,endfloats]{revtex4}
\usepackage{graphicx}
\usepackage{mathrsfs}
\usepackage{wasysym}
\usepackage{textcomp}
\usepackage{pifont}

\begin{document}

\title{Resonant transparency of materials with negative permittivity}
\author{E. Fourkal$^{1}$}
\author{I. Velchev$^{1}$}
\author{C-M. Ma$^{1}$}
\author{A. Smolyakov$^{2}$}

\affiliation{$^{1}$Department of Radiation Physics, Fox Chase Cancer Center,
Philadelphia, PA 19111, U.S.A.} 
\affiliation{$^{2}$Department of Physics and
Engineering Physics, University of Saskatchewan, Saskatoon, Canada}
%%%%%%%%%%%%%%%%%%%%%%%%%%%%%%%%%%%%%%%%%%%%%%%%%%%%%%%%%%%%%%%%%%%%%%%%%%%%%%%%%%%%%%%%%%%%%%%%%%%%%%%%%%%%%%%%%%%%%%%%%%%%
%\usepackage{amssymb,graphicx}

%\journal{Physics Letters A}

%\begin{document}

%\begin{frontmatter}
%\title{Resonant transparency of materials with negative permittivity}
%\author[FCCC]{E. Fourkal\corauthref{cor}},
%\corauth[cor]{Corresponding author.}
%\ead{eugene.fourkal@fccc.edu}
%\author[FCCC]{I. Velchev},  
%\author[FCCC]{C-M. Ma},
%\author[Usask]{A. Smolyakov}

%\address[FCCC]{Department of Radiation Physics, Fox Chase Cancer Center,
%Philadelphia, PA 19111, U.S.A.}
%\address[Usask]{Department of Physics and
%Engineering Physics, University of Saskatchewan, Saskatoon, Canada}

\begin {abstract}
It is shown that the transparency of opaque material with negative permittivity
exhibits resonant behavior. The resonance occurs as a result of the excitation of the surface waves at slab
boundaries. Dramatic
field amplification of the incident evanescent fields at the resonance improves the resolution of the
the sub-wavelength imaging system (superlens). 
A finite thickness slab can be totally
transparent to a \textit{p}-polarized obliquely incident electromagnetic
wave for certain values of the incidence angle and wave frequency corresponding to the excitation of the surface modes.
At the resonance,  two evanescent waves have a finite phase shift
providing  non-zero energy flux through the non-transparent region. 
\end {abstract}

\date{\today}
%\begin{keyword}
%Overdense plasma, surface modes, evanescent wave interference, superlensing
%\PACS 52.25.Os \sep 42.30.Lr \sep 52.35.Lv \sep 42.25.Hz
%\end{keyword}
%\end{frontmatter}

\maketitle

\section{Introduction}

Propagation of the electromagnetic radiation in materials with negative
dielectric permittivity (permeability) or the so called left-handed
materials (LHM) has attracted great deal of attention in recent years~\cite
{kindel75,dragila85,bliokh05,fang05}. The increased interest in properties
of such media has been driven by their potential applications in various
branches of science and technology. One possible application is related to
the possibility of creating the so called superlens: a subwavelength
optical imaging system without the diffraction limit~\cite
{fang05,shvets04,pendry00,podolskiy05}. The superlens phenomenon is
essentially based on amplification of evanescent waves, facilitated by the
excitation of the surface plasmons \cite{pendry00}. Plasma with overcritical
density is a simplest example of the negative $\epsilon$ material; $\epsilon
=1-\omega _p^2/\omega ^2<0$ for $\omega <\omega _p$. Phenomena that take
place in such plasmas are important in a number of areas, in particular for
the inertial confinement fusion (ICF) experiments~\cite
{kindel75,aliev77,bychenkov01}.

In this work we show that the amplification of the evanescent waves
originating from the interaction of two evanescent fields (decaying and
growing in space) has a resonant character related to the excitation of
surface modes. Such amplification of the evanescent field allows the
penetration of the electromagnetic radiation to depths much greater compared
to the incident light wave length\cite{dragila85}. The evanescent wave
incident on the negative $\epsilon$ slab is strongly amplified when the
resonant conditions for the surface modes are met. For materials with $%
\epsilon \neq -1$ the resonance occurs for finite values of the wave vector $%
k_y$ (for $\epsilon=-1$, the resonance occurs for $|k_y| \rightarrow \infty)$%
, where $k_y$ is the in-plane wave vector component (p-polarization is
considered). The presence of resonances with finite values of the in-plane
wave vector $k_y$ may significantly improve the overall resolution of the
subwavelength imaging system.

Resonant excitation of a surface mode is also the underlying mechanism
behind the absolute transparency of a finite thickness slab of material with 
$\epsilon <0$ to the incident propagating electromagnetic waves ($%
|k_y|<\omega /c$)\cite{dragila85,fourkal06}. In this case, the resonant
excitation of surface modes by the incident light can be achieved via the
presence of a single transition layer with $0<\epsilon <1$ on one side of
the slab with $\epsilon <0$\cite{fourkal06}. The superposition of two
evanescent waves provides a finite energy flux through the region with $%
\epsilon <0$, which is equal to that in the incident electromagnetic wave.
The radiation is then re-emitted at the other side of the opaque slab.

\section{Interference of the evanescent waves and effect of superlensing}

Non-propagating (evanescent) modes are basic solutions to the Maxwell's
equations for the electromagnetic fields in materials with negative
dielectric permittivity. Often such modes have been neglected assuming that
the boundary conditions in an infinite medium preclude the exponentially
growing modes while the decaying modes do not contribute to the
transmission. It has recently been noted however \cite{pendry00} that in a
slab of material with negative permittivity both (growing and decaying)
components are present, resulting in amplification of the evanescent modes.
As shown by Pendry~\cite{pendry00}, the effect of amplification of the
evanescent spectrum of the incident light (originating from the object) by
"negative" materials can be advantageously used for subwavelength imaging
applications, potentially leading to optical system without the diffraction
limit or superlens.

\begin{figure}[t]
\centering
\centerline{\includegraphics[width=0.9\columnwidth]{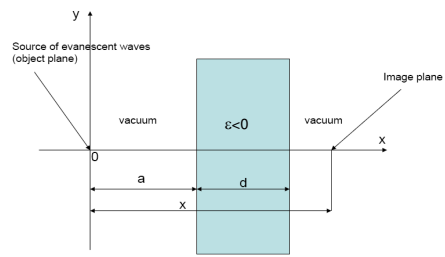}}
\caption{Schematic geometry of the dielectric constant distribution for the
case of a single planar medium.}
\label{fig1}
\end{figure}

\begin{figure}[t]
\centering
\centerline{\includegraphics[width=0.9\columnwidth]{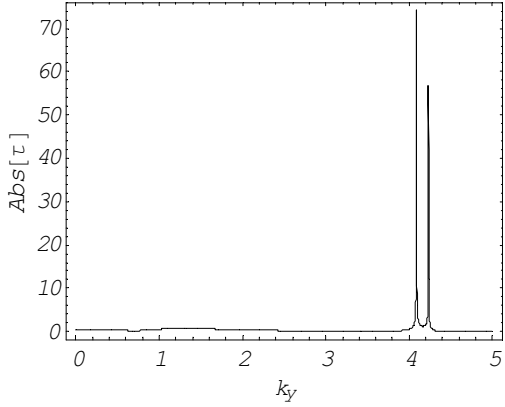}}
\caption{The absolute value of the optical transfer function at a distance $%
x=2d$ from the source as a function of the in-plane wave vector $k_y$. The
dielectric constant of the medium and its thickness are $%
\epsilon=-1.0292+0.00001i$ and $d=1.8c/\omega_p$ correspondingly. $k_y$ is
normalized to classical skin depth $c/\omega_p$. External radiation
frequency $k_0=0.702$.}
\label{fig2}
\end{figure}

The imaging problem can be described in terms of the optical transfer
function $\tau (x,k_y,\omega )$ ($k_y$ designates the in-plane wave vector
directed along the surface of the material), defined as the ratio of Fourier
components of image field to object field, $%
B_{img}^{k_y}(x)/B_{obj}^{k_y}(0) $ (for $-\infty <k_y<\infty $) at a given
imaging plane $x$. The transfer function can be found by considering a $p$
-polarized wave (electric vector in the plane of incidence) incident from
vacuum on a thin overcritical density slab of thickness $d$, dielectric
permittivity $\epsilon <0$ and magnetic permeability $\mu =1$ as shown in
Figure~\ref{fig1}. The optical properties of this slab are obtained by
taking the ratio of the field in the region $x>a+d$ to that at the object
plane (in current calculations the object plane is assumed to be at $x=0$).
The electromagnetic fields in each region of interest are found from solving
the well known wave equation, 
\begin{equation}
\epsilon \frac d{dx}\left( \frac 1\epsilon \frac{dB_z}{dx}\right) +\frac{%
\omega ^2}{c^2}\left( \epsilon -\frac{k_y^2c^2}{\omega ^2}\right) B_z=0,
\label{eq2}
\end{equation}
with a general solution having the following form, 
\begin{equation}
B_z=\left( A_1e^{ikx}+A_2e^{-ikx}\right) e^{i(k_yy-\omega t)}  \label{eq2a}
\end{equation}
where $k=\omega /c\sqrt{(\epsilon -k_y^2c^2/\omega ^2)}$. The
electromagnetic fields in vacuum regions ($x<a$ and $x>a+d$) represent a sum
of incident and reflected waves ($x<a$), and a transmitted wave ($x>a+d$).
Matching solutions at different boundaries by requiring the continuity of $%
B_z$ and $1/\epsilon dB_z/dx$ across interfaces, one arrives at the
expression for the transfer function, 
\begin{eqnarray}
&&\tau (x,k_y,k_0)=\frac{2k_0^2\epsilon \sqrt{\frac{k_y^2}{k_0^2}-1}\sqrt{%
\epsilon -\frac{k_y^2}{k_0^2}}}{\Xi +\Lambda }e^{k_0\sqrt{\frac{k_y^2}{k_0^2}%
-1}(d-x)}  \label{eq4} \\
&&\Xi (d,k_y,k_0)=2k_0^2\epsilon \sqrt{\frac{k_y^2}{k_0^2}-1}\sqrt{\epsilon -%
\frac{k_y^2}{k_0^2}}\cos [k_0d\sqrt{\epsilon -\frac{k_y^2}{k_0^2}}] 
\nonumber \\
&&\Lambda (d,k_y,k_0)=\left( \left( 1+\epsilon ^2\right) k_y^2-\epsilon
\left( 1+\epsilon \right) k_0^2\right) \sin [k_0d\sqrt{\epsilon -\frac{k_y^2%
}{k_0^2}}],  \nonumber
\end{eqnarray}
where $x$ is a distance between the source and the image planes and $%
k_0=\omega /c$. Figure~\ref{fig2} shows the absolute value of the optical
transfer function at a distance $x=2d$ from the source. As one can see,
there are in general two sharp peaks in the transfer function occurring for
certain resonant values of the in-plane wave vector $k_y>\omega /c$ where
the spectrum is completely evanescent. The peaks correspond to the resonant
excitation of surface plasmons that are supported by the interface between
the thin overcritical density medium and vacuum. This means that only those
fields for which $k_y>\omega /c$ may resonantly couple to this particular
surface mode and get dramatically amplified through constructive
interference (these are the spectral components that carry
sub-diffraction-limited resolution of the object). In fact, the zeros of the
denominator in Eq.~\ref{eq4} 
\begin{equation}
\Xi (d,k_y,k_0)+\Lambda (d,k_y,k_0)=0  \nonumber
\end{equation}
define the dispersion relation for the plasma surface eigenmode that is
supported by the thin overdense medium. The above equation can be reduced to
the following form: 
\begin{equation}
\tanh [\sqrt{k_y^2-\epsilon k_0^2}d]=\frac{2\epsilon \sqrt{k_0^2-k_y^2}\sqrt{%
\epsilon k_0^2-k_y^2}}{(1+\epsilon ^2)k_y^2-\epsilon (1+\epsilon )k_0^2}
\label{eq5}
\end{equation}
In the general case, for a finite value of $d$ there exist two solutions to
Eq.~\ref{eq5}, as seen from Figure~(\ref{fig2}) corresponding to a coupled
surface wave running on opposite sides of the slab. As $d$ becomes large so
that $\tanh [\sqrt{k_y^2-\epsilon k_0^2}d]\rightarrow 1$, the two solutions
degenerate and we obtain, 
\begin{equation}
k_y=k_0\sqrt{\frac \epsilon {1+\epsilon }},  \label{eq6}
\end{equation}
which is the well known dispersion relation for surface plasmons on an
overdense medium-vacuum interface. It is worth noting here that the transfer
function given by expression~\ref{eq4} is exact and no approximations were
used in its derivation. It includes the resonant contribution of surface
modes with a finite in-plane wave vector $k_y$ and in the limit $\epsilon
\rightarrow -1$, $|k_y|>>\omega /c$ it is reduced to that obtained in Ref.~%
\cite{podolskiy05} where the authors investigated the resolution limit of
the slab of ''negative'' material with dielectric permittivity $\epsilon
=-1+i\epsilon ^{^{\prime \prime }}$ and magnetic permeability $\mu =-1+i\mu
^{^{\prime \prime }}$. It should also be noted that the location (in $k_y$
wave vector space) of the two resonances not only depends on the absolute
value of the dielectric constant $\epsilon$ but also the thickness of the
slab $d$. For a very thin slab, the two resonances are separated (smaller slab
thicknesses result in greater peak separation in $k_y$ space). As the slab
thickness increases, the two resonant points move toward each other,
eventually merging together forming a single peak. All of these features in
the transfer function have not been noticed in the previous investigations
since the authors limited their calculations to the asymptotic case $%
k_y>>\omega/c$ and $k_yd>1$. As we will see below, the presence of such
resonances may advantageously be used in the design of superlens for
improved image resolution and intensity.

\begin{figure}[t]
\centering
\centerline{\includegraphics[width=0.9\columnwidth]{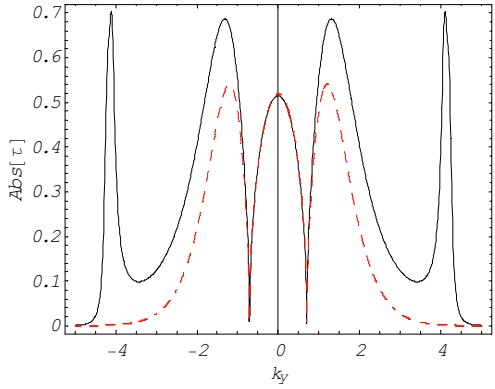}}
\caption{The absolute value of the optical transfer function at a distance $%
x=2d$ from the source as a function of the in-plane wave vector $k_y$ for
two different materials with dielectric permittivities $%
\epsilon_1=-1.0292+0.001i$ (solid line) and $\epsilon_2=-1+0.001i$ (dashed
line) and thickness $d=1.8c/\omega_p$. $k_y$ is normalized to classical skin
depth $c/\omega_p$}
\label{fig3}
\end{figure}

The transfer function $\tau $ can be used to find the reconstructed field in
the image plane in the form 
\begin{equation}
B_{img}(x,y,t)=\int S(k_y)\tau (x,k_y,k_0)e^{i(k_yy-\omega t)}dk_y,
\label{eq7}
\end{equation}
where $S(k_y)$ is the wave vector spectrum of the source (imaged object).
Thus, the ability of the system to image the object is completely determined
by the optical transfer function~(\ref{eq4}), which in itself depends on
many system's physical parameters. In an ideal case, the transfer function
should transfer all spatial harmonics equally or $\tau (x,k_y,k_0)=1$ for $%
-\infty <k_y<\infty $. In reality however, the transfer function is a
non-monotonous function of the wave vector $k_y$, medium material type
defined by $\epsilon $, its thickness $d$, and the position $x$ of the
imaging plane relative to the position of the object plane. The superlensing
can only be realized for certain set of the above parameters. Figure~\ref
{fig3} shows the optical transfer function for two layers of materials with
dielectric permittivities $\epsilon _1=-1+0.001i$ and $\epsilon
_2=-1.0292+0.001i$ at the imaging plane $x=2d$. As one can see the
transmission band (in the wave vector space) is significantly broadened in
the second material ($\epsilon =-1.0292+0.001i$) due to the resonant
excitation of a surface plasmon with a finite value of the wave vector $%
k_y\approx 4$. Such surface plasma wave is not excited in material with $%
\epsilon =-1$ (assuming the slab thickness $d$ such that the condition $%
\tanh [\sqrt{k_y^2-\epsilon k_0^2}d]=1$ is satisfied), as can be seen from
the dispersion relation (\ref{eq6}).

\begin{figure}[t]
\centering
\centerline{\includegraphics[width=0.9\columnwidth]{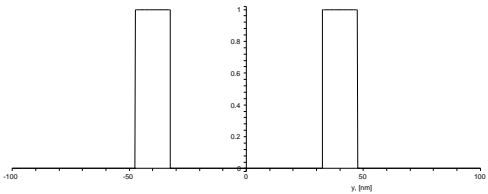}}
\caption{The field distribution in the object plane. The object comprises
two slabs of thickness 15 nm, separated by a distance 80 nm from their
centers.}
\label{fig4}
\end{figure}

\begin{figure}[t]
\centering
\centerline{\includegraphics[width=0.9\columnwidth]{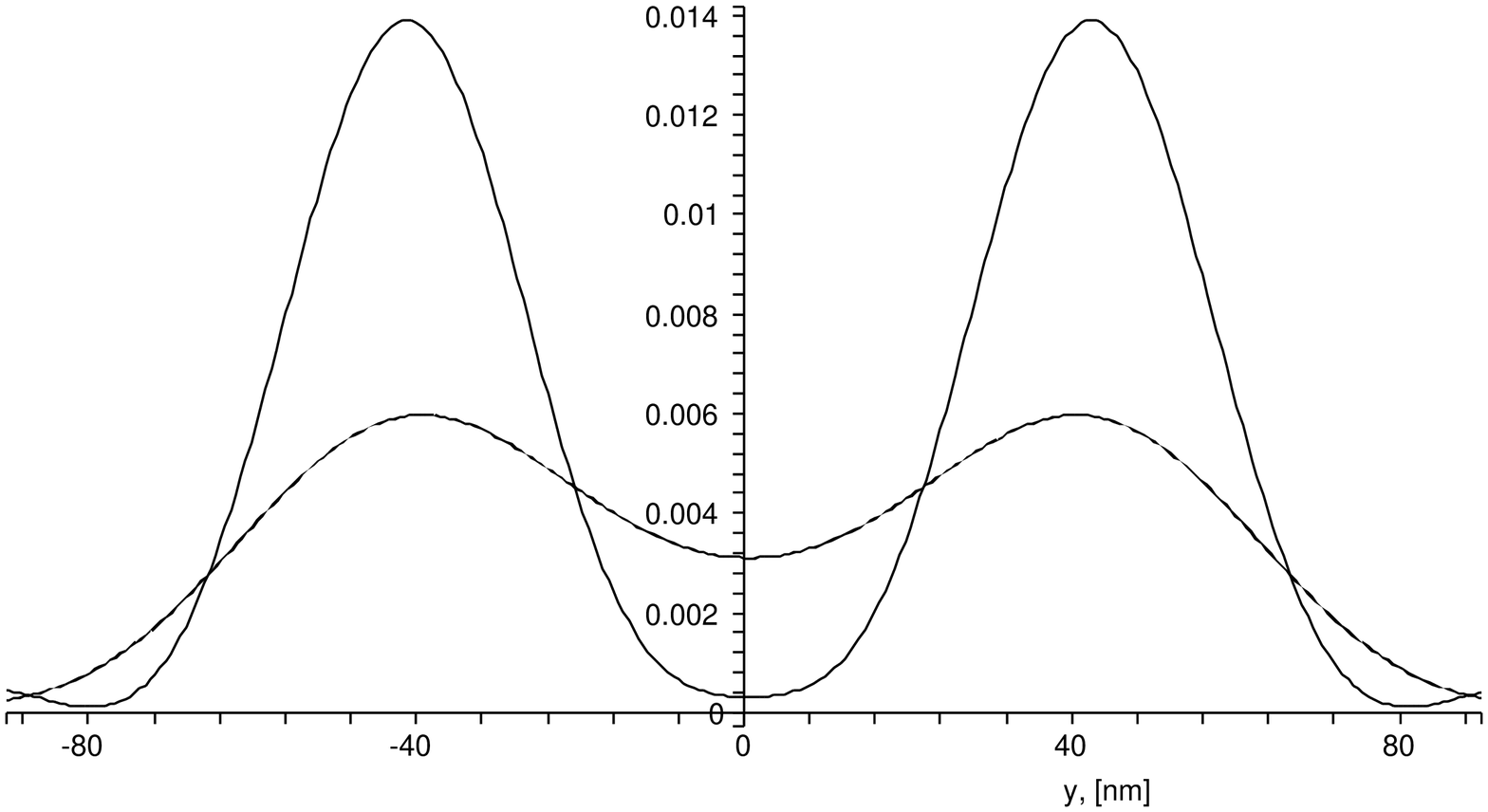}}
\caption{Field distribution in the image plane for two slabs with $%
\epsilon_1=-1.0292+0.001i$ (solid line) and $\epsilon_2=-1+0.001i$ (dotted
line) and thickness $d=1.8c/\omega_p$. Magnetic permeability of both
materials is assumed to be $\mu=1$. The wavelength of the incident light is $%
\lambda=350$ nm. }
\label{fig5}
\end{figure}

Let us estimate how well we can image an object using these two materials.
For a sake of simplicity we assume that our object is represented by two
slits of a certain width located at a given distance away from each other as
shown in Figure~\ref{fig4}. Substituting the Fourier transform of the object
together with the transfer function~(\ref{eq4}) into Eq.~\ref{eq7} we arrive
at the reconstructed image shown in Figure~\ref{fig5} for the case when the
incident light wavelength is $\lambda=350$ nm. As one can see, both
materials provided considerable focusing, yielding the sub-wavelength
resolution of the object. However, as we expected the image resolution and
its intensity for the second layer ($\epsilon=-1.0292$) is superior to that
using material with $\epsilon=-1$. This suggests that there is a range of
system parameters for which a significant focusing can be achieved\cite
{fang05} necessitating a further parametric study in order to understand the
relation between different physical parameters of the system.

\section{Surface wave induced total transparency of material with negative
permittivity}

Surface wave induced amplification of incident electromagnetic waves with
spectral components of in-plane wave vectors $k_y$ satisfying the condition $|k_y|>\omega/c$ has been considered in the previous section. It was shown that
the resonant amplification occurs as a result of the excitation of a surface
wave. A slab of negative $\epsilon $ material surrounded by vacuum supports
such a surface mode for which its phase velocity is always sub-luminal or $
\omega /k_y<c$. As a result, only those modes of the incident light for
which $|k_y|>\omega /c$ have been amplified. It is possible however to amplify the
propagating modes with $|k_y|<\omega /c$ too, thus creating the conditions for
the absolute transparency to the incident propagating wave. This can be done
by creating conditions for the excitation of a surface mode with phase
velocity greater than that of the speed of light.

\begin{figure}[t]
\centering
\centerline{\includegraphics[width=0.9\columnwidth]{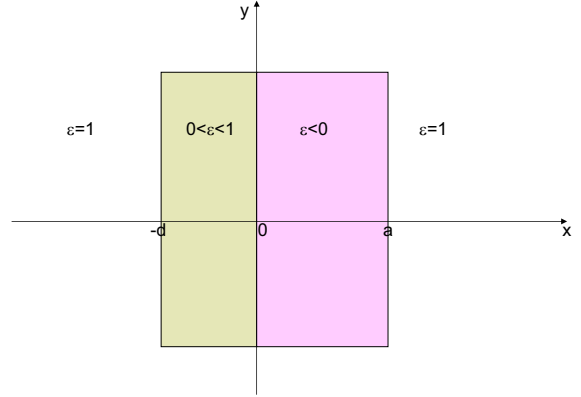}}
\caption{Schematic diagram of the spatial density (dielectric constant)
distribution for the case of a layered slab.}
\label{fig6}
\end{figure}

\begin{figure}[t]
\centering
\centerline{\includegraphics[width=1.2\columnwidth]{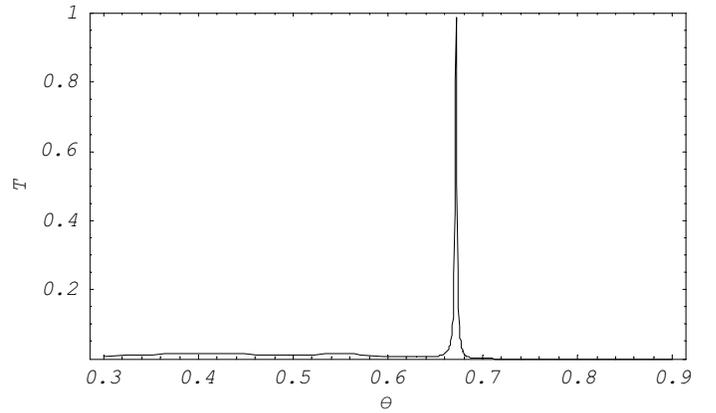}}
\caption{Transmission coefficient as a function of the incidence angle. A
double-layer plasma system becomes completely transparent at the incidence
angle $\theta$=0.671955. Undercritical and overcritical plasma slabs
dielectric constants and thicknesses are $\epsilon_1=0.3428$, $%
d=27*c/\omega_{p2}$ and $\epsilon_2$=-2.97, $a=3.12*c/\omega_{p2}$
correspondingly. External radiation frequency $\omega/\omega_{p2}=0.5019$.}
\label{fig7}
\end{figure}

Consider $p$-polarized light obliquely incident from vacuum on a two-layer
structure having dielectric permittivity distribution shown in Fig.~\ref
{fig6}. Such a system can be formed by placing an undercritical density
plasma layer (with thickness $d$ and electron density corresponding to the
plasma frequency $\omega _{p,1}$) to an overcritical density plasma layer
(with thickness $a$ and electron density corresponding to the plasma
frequency $\omega _{p,2}$). The plasma-plasma interface supports a surface
wave with dispersion relation found from the solution to the Maxwell's
equations\cite{kindel75}, 
\begin{equation}
\alpha _1/\epsilon _1+\alpha _2/\epsilon _2=0\rightarrow k_y=\frac{\sqrt{%
\left( \Delta -\omega ^2\right) \left( \omega ^2-1\right) }}{\sqrt{1+\Delta
-2\omega ^2}},  \label{eq1}
\end{equation}
where $\alpha _i^2=k_y^2-\epsilon _i\omega ^2/c^2$, $\epsilon _i=1-\omega
_{p,i}^2/\omega ^2$, $(i=1,2)$; $\Delta =\omega _{p,1}^2/\omega _{p,2}^2$; $%
\omega $ is normalized to the plasma frequency in the region where $\epsilon
<0$ and $k_y$ is normalized to the classical skin depth $\delta =c/\omega
_{p,2}$. It can be easily seen that the phase velocity of the surface wave
on a plasma-plasma interface can be greater than that of light, so that they
can couple to radiating electromagnetic fields. This means that the incident 
$p$-polarized electromagnetic wave may excite a surface mode on a
plasma-plasma interface if the resonant condition (external field frequency
and its wave vector's tangential component have to match those that are
determined from the dispersion relation for the plasma surface wave) is
satisfied. The optical properties of this dual-layer system can be found
from matching the fields of the incident/reflected electromagnetic waves
with those of the surface wave at the three interfaces. The electromagnetic
field in each region $\mathbf{B}=(0,0,B_z)$ is a solution to the wave
equation \ref{eq2} with general solution given by expression~\ref{eq2a}. The
electromagnetic fields in vacuum regions ($x<-d$ and $x>a$) represent a sum
of incident and reflected wave ($x<-d$), and a transmitted wave ($x>a$).
Matching solutions at different boundaries by requiring continuity of $B_z$
and $1/\epsilon dB_z/dx$ across interfaces, we obtain the unknown expansion
coefficients with the transmission coefficient having the following form, 
\begin{eqnarray}
&&T=\left| \frac{4e^{i(k_1d+k_2a-(a+d)k_0\cos \theta )}k_0k_1k_2\epsilon
_1\epsilon _2\cos \theta }{\left( \Re -\Upsilon \right) }\right| ^2
\label{eq3} \\
&&\Re =\left( 1+e^{2ik_1d}\right) k_0k_1k_2\epsilon _1\epsilon _2\cos \theta
+e^{2ik_2a}\left( 1+e^{2ik_1d}\right)  \nonumber \\
&&k_0k_1k_2\epsilon _1\epsilon _2\cos \theta  \nonumber \\
&&\Upsilon =2ie^{i(k_1d+k_2a)}k_2\epsilon _2\cos [k_2a]\left(
k_1^2+k_0^2\epsilon _1^2\cos ^2\theta \right) \sin [k_1d]  \nonumber \\
&&+k_1\epsilon _1\cos [k_1d]\left( k_2^2+k_0^2\epsilon _2^2\cos ^2\theta
\right) \sin [k_2a]  \nonumber \\
&&-ik_0\left( k_2^2\epsilon _1^2+k_1^2\epsilon _2^2\right) \cos \theta \sin
[k_1d]\sin [k_2a].  \nonumber
\end{eqnarray}
where $\theta =\arcsin [k_y/k_0]$. Figure~\ref{fig7} shows the transmission
coefficient as a function of the incidence angle. As one can see there is a
sharp increase in the transmission properties of the system when the angle
of incidence matches certain resonant value of $\theta $ at which point the
transmission coefficient reaches unity. The resonant value exactly
corresponds to that given by an expression (\ref{eq1}) for the dispersion
relation of the surface plasma wave. Thus, the anomalous transmission occurs
even for a system consisting of an undercritical density plasma layer
adjacent to an overcritical density plasma slab, so that there is no need to
form a sandwich-like structure as argued in Refs.~\cite{dragila85,dragila87}
in which a second surface plasma wave has to be excited on the opposite side
of the overcritical density layer to achieve the same effect. In other
words, the anomalous light transmission can be achieved through excitation
of a single surface plasma wave. The transmission coefficient is also a
non-monotonous function of both plasma layer thickness with a single maximum
reached at certain correlated values determined from the interference
condition between evanescent fields in both plasma slabs.

\section{Evanescent wave interference and the energy transport}

It is often assumed that the evanescent waves do not carry the energy.
Therefore, in the problem of total transparency of a layered structure there
occurs a question of how the energy is carried through the non-transparent
media where the only solutions are the evanescent modes. One must remember
however that the general solution inside the negative $\epsilon $ medium is
a sum of two exponential functions, one that decays with the distance $\sim
e^{-x}$ ($x$ points in the propagation direction) and the other grows $\sim
e^x$. For the superposition of decaying and growing modes, $E,B\sim A_1\exp
[-ikx]+A_2\exp [ikx]$ ($k$ is purely imaginary decay constant), the $x$
component of the time averaged Poynting vector $S_x$  
\begin{equation}
S_x=\frac 12Re[E_yB_z^{*}]\sim Re[k(A_1A_2^{*}-A_2A_1^{*})] \sim
2Im[A_1A_2^{*}],  \nonumber
\end{equation}
may become finite when the combination $A_1A_2^{*}$ has a finite imaginary
part, which  requires a finite phase shift between $A_1$ and $A_2$. 
Therefore, a finite energy flux occurs as a result of the superposition of
two evanescent modes with a finite phase shift . We shall call this the
interference of the evanescent modes.

It is easy to show that the required phase shift can be obtained when two
evanescent modes inside the negative $\epsilon$ region are matched to
the outgoing (transmitted) wave in vacuum. Matching of the evanescent
solutions with the incident vacuum wave (at the other side of the negative $
\epsilon$ region) shows that the total transmission may be obtained only
when the transition layer with $\epsilon > 0$ is included\cite
{fourkal06}. The condition of the absolute transparency is equivalent to
the resonant condition for the excitation of the surface plasma mode. At the
resonance the Poynting flux inside the slab becomes equal to that of the
incident radiation and the opaque plasma slab becomes absolutely
transparent. This can be a possible  mechanism of the anomalously high
transparency of overdense plasma recently observed in the experiment\cite
{bychenkov05}. 

\section{Summary}

In conclusion, we have shown that the excitation of surface modes leads to
the resonant transparency of optically opaque materials. Presence of such
resonances may improve the resolution and the signal intensity of the sub
wavelength imaging system. It has been also shown that the resonant
excitation of surface modes is an underlying mechanism behind the total
transparency of an overdence plasma slab to the incident electromagnetic
wave. The energy flux through the negative $\epsilon $ region occurs as a
result of the interference of the evanescent modes.

This work is in part supported by Strawbridge Family Foundation, and NSERC
Canada.

.

\end{document}